\documentclass{jfm}

\usepackage{amsmath}
\usepackage{color}
\usepackage{xfrac}

\usepackage{graphicx}
\usepackage{amssymb}
\usepackage{amsbsy}
\usepackage{upmath}
\usepackage{textcomp}
\usepackage{natbib}

\newcommand{\RomanNumeralCaps}[1]
\linenumbers

\shorttitle{Bubble clouds in viscoplastic fluids}
\shortauthor{E.~Chaparian \& I.~A.~Frigaard}

\title{Clouds of bubbles in viscoplastic fluids}

\author{Emad Chaparian\aff{1}
\and
Ian A.~Frigaard\aff{1,2}
\corresp{\email{emad@math.ubc.ca}}}

\affiliation{
\aff{1}Department of Mathematics, University of British Columbia, 1984 Mathematics Road, Vancouver, BC, Canada, V6T 1Z2
\aff{2}Department of Mechanical Engineering, University of British Columbia, 6250 Applied Science Lane, Vancouver, BC, Canada, V6T 1Z4
}

\begin{document}
\maketitle

\begin{abstract}
Viscoplastic fluids can hold bubbles/particles stationary by balancing the buoyancy stress with the yield stress. In the present study, we investigate a suspension of bubbles in a yield-stress fluid. More precisely, we compute how much the gas fraction $\phi$ that could be held trapped in a yield-stress fluid without motion. The key parameter is the yield number which represents the ratio of the yield stress to the buoyancy stress. Here the goal is to shed light on how the bubbles feel their neighbours through the stress field and to compute the critical yield number for a bubble cloud beyond which the flow is suppressed. We perform 2D computations in a full periodic box with randomized positions of the monosized circular bubbles. A large number of configurations are investigated to obtain statistically converged results. We intuitively expect that for higher volume fractions the critical yield number is larger. Not only here do we establish that this is the case, but also we show that short range interactions of bubbles increase the critical yield number even more dramatically for bubble clouds. The results show that the critical yield number is a linear function of volume fraction in the dilute regime. An algebraic expression  model is given to approximate the critical yield number (semi-empirically) based on the numerical experiment in the studied range of $0\le \phi \le 0.31$, together with lower and upper estimates.
\end{abstract}


\section{Introduction}
\label{sec:intro}

Bubbles in yield-stress fluids arise in both geophysical and industrial processes, ranging from bubbling mud pits through aerated chocolate to foamed cement. This fact has stimulated a number of studies, on both individual bubbles and multiple bubbles. The latter has mostly concentrated on the rheological behaviour of these mixtures and especially foamy yield-stress fluids. \cite{kogan2013mixtures} generalized a theoretical homogenization framework introduced initially for suspensions of particles in yield-stress fluids \citep{chateau2008homogenization} and  studied the shear rheology of these materials experimentally. \cite{goyon2010shear} also investigated the drainage of foamy materials induced by shear.

In this study however, we focus on the stability of a cloud of bubbles in a yield-stress fluid which is directly relevant to a large number of applications in which the mixture remains stationary. The oil and gas industry has long used foamed cements (and drilling fluids) in well construction  \citep{Benge1982,Ahmed2009}. Major themes of the investigation of the Deepwater Horizon oil spill \citep{Macondo2011} concerned the stability of the foamed slurry downhole, its testing and suitability for this well. In the wider construction industry, both escaping and trapped bubbles can be desirable in cement pastes, either entrained into the slurry during processing or purposefully foamed. Producing an air void system within concrete by inducing rising bubbles helps concrete to become resistant to freeze-thaw cycles, and thus bubble rise in fresh cement paste is of interest \citep{ley2009observations}.

Our motivation comes from a different process: gas emissions from tailings ponds resulting from oil sand production. In these ponds, fine and mature fluid tailings form stratified layers which do not appear to consolidate significantly over timescales of many decades. The bulk rheology of this layer exhibits a yield stress \citep{derakhshandeh2016kaolinite}. Anaerobic microorganisms bio-degrade naptha producing methane, which can be one of the main sources of gas emission from tailing ponds. Carbon dioxide is also produced \citep{Small2015}. In this case, the ideal scenario will be to prevent bubbles from rising or indeed we might wish to estimate what is a ``safe'' trapped gas fraction to be held in the pond. Similar mechanisms in geological materials, such as shallow marine, terrestrial sediments and in some flooded soils, also lead to the formation bubbles \citep{Boudreau2012}.

Motion of an individual bubble in a yield-stress medium has been studied many times with different approaches. Here we use the simplest viscoplastic model, i.e.~Bingham fluid, since we are interested in the onset of motion, which is the same for any ``simple'' yield-stress fluid model \citep{frigaard2019background}. Tsamopoulos and co-workers \citep{tsamopoulos2008steady,dimakopoulos2013steady} in a series of papers investigated this problem using different numerical schemes and reported drag coefficients and steady shapes of bubbles for a wide range of effective parameters such as the Reynolds, Bingham and Bond numbers. Experimental studies \citep{sikorski2009motion,lopez2018,pourzahedi2021experimental} have explored the velocity and shape of air bubbles rising through Carbopol gel, where elasticity of the yield-stress fluid causes a fore-aft asymmetry in the bubble shapes (a tear-drop shape). Some analytical models have been developed to capture this phenomenon \citep{sun2020dynamic}.

Nevertheless, in the subject of the present study, there is little direct numerical/experimental work to describe the onset of motion. The motion onset problem was first formulated mathematically by \cite{Dubash2004}. Very recently, we conducted a systematic study on the yielding of an individual bubble with different shapes and surface tensions \citep{pourzahedi2021}. Meanwhile, \cite{chaparian2018inline} have demonstrated that a cluster of particles (with bridges of unyielded material which connect the particles together) can be formed when particles are close enough in a yield-stress fluid, which dramatically increases the critical yield number. \cite{koblitz2018DNS} have reported the same phenomenon on investigating sedimentation limits in a dilute suspension of rigid particles within a yield-stress fluid.

Here we focus on a cloud of bubbles and how the bubbles feel their neighbours and interact with each other. We compute the critical yield number for a bubble cloud beyond which the flow is suppressed and explore the different contributing influences. An outline of the paper is as follows. In \S \ref{sec:formulation}, we set out the problem and review the key features of the implemented numerical method. The main results are presented in \S \ref{sec:results} and conclusions drawn in \S \ref{sec:conclusion}.

\section{Problem statement}
\label{sec:formulation}

\subsection{Mathematical formulation}

We consider inertialess incompressible bubbly flow of a yield-stress fluid governed by the non-dimensional equation,
\begin{equation}\label{eq:non-govern}
0 = - \boldsymbol{\nabla} p + \boldsymbol{\nabla} \boldsymbol{\cdot} \ubtau - \frac{1}{1-\rho} \boldsymbol{e}_g,~~\text{in} ~\Omega \setminus \bar{X},
\end{equation}
and the Bingham model,
\begin{equation}\label{eq:non-const}
  \left\{
    \begin{array}{ll}
      \ubtau = \left( 1 + \displaystyle{\frac{Y}{\Vert\dot{\ubgamma}\Vert}} \right) \dot{\ubgamma} & \mbox{iff}\quad \Vert \ubtau \Vert > Y, \\[2pt]
      \dot{\ubgamma} = 0 & \mbox{iff}\quad \Vert \ubtau \Vert \leqslant Y .
  \end{array} \right.
\end{equation}
Here $p$ is the pressure inside the ambient yield-stress liquid, $\ubtau$ the deviatoric stress tensor, $\rho = \hat{\rho}_b / \hat{\rho}_l$ the ratio of the bubble density to the liquid density, $\boldsymbol{e}_g$ the basis vector in the gravity direction and $Y = \hat{\tau}_y / \Delta \hat{\rho} \hat{g} \hat{R}$ is the yield number ($\Delta \hat{\rho} = \hat{\rho}_l - \hat{\rho}_b$). Here, We scaled the dimensional pressure ($\hat{p}$) and the deviatoric stress tensor ($\hat{\ubtau}$) with the buoyancy stress $\left( \hat{\rho}_l - \hat{\rho}_b \right) \hat{g} \hat{R}$ and the velocity vector ($\hat{\boldsymbol{u}}=(\hat{u},\hat{v})$) with the velocity,
\[
\hat{U} = \frac{\Delta \hat{\rho}  \hat{g} \hat{R}^2}{\hat{\mu}_l},
\]
which arises from balancing the buoyancy stress with a characteristic viscous stress ($\hat{\mu}_l \hat{U}/\hat{R}$); here $\hat{R}$ is the radius of the monodispersed circular bubbles and $\hat{\mu}_l$ the plastic viscosity of the liquid. Quantities with the hat symbol ($\hat{\cdot}$) are dimensional. In (\ref{eq:non-const}),  $\dot{\ubgamma}$ is  the  rate of strain  tensor  and $\Vert \cdot \Vert$ is  the norm associated with the tensor inner product:
\[
\boldsymbol{c} \boldsymbol{:} \boldsymbol{d} =\frac{1}{2} \sum_{ij} c_{ij} ~d_{ij};
\]
e.g.~$\Vert \ubtau \Vert = \sqrt{\ubtau \boldsymbol{:} \ubtau}$. Note that generally for bubbles $\hat{\rho}_b \ll \hat{\rho}_l$, hence in practice $\Delta \hat{\rho} \approx \hat{\rho}_l$ and $\rho \approx 0$. The full flow domain (yield-stress fluid and bubbles) is denote by $\Omega$, the gas fraction by $X$ and the bubble surfaces by $\partial X$.  Hence the bubble area fraction is $\phi=\text{meas}(X)/\text{meas}(\Omega)$.

On the bubble surfaces ($\partial X$) the jump in the straction vector is balanced by the surface tension in the normal direction. In the inviscid limit ($\hat{\mu}_b \approx 0$), the tangential stress vanishes: \begin{equation}\label{eq:YL_tang}
\left( \ubsigma \boldsymbol{\cdot} \boldsymbol{n} \right) \boldsymbol{\cdot} \boldsymbol{t} = 0,~~~\mbox{on} ~\partial X,
\end{equation}
and the normal component satisfies
\begin{equation}\label{eq:YL_normal}
-p+p_b+(\ubtau \boldsymbol{\cdot} \boldsymbol{n}) \boldsymbol{\cdot} \boldsymbol{n} = \frac{\gamma}{\kappa},~~~\mbox{on} ~\partial X,
\end{equation}
where $p_b$ is the pressure inside the bubble, $\kappa (=1)$ is the radius of curvature and $\gamma~(= \hat{\gamma} / \Delta \hat{\rho} \hat{g} \hat{R}^2  )$; $\hat{\gamma}$ is the surface tension coefficient.

\begin{figure}
\centerline{\includegraphics[width=0.7\linewidth]{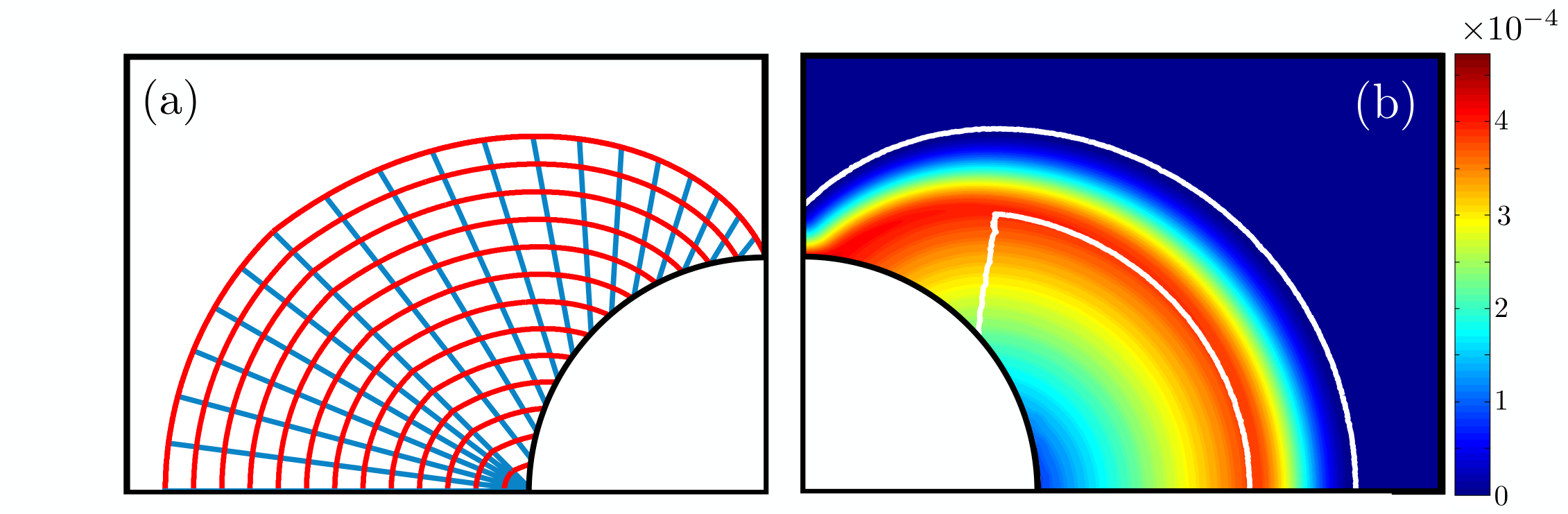}}
\caption{(a) Slipline solution about a circular bubble (see \cite{chaparian2020sliding} and \cite{pourzahedi2021} for more details). (b) Speed contour $\vert \boldsymbol{u} \vert$ at $Y=0.17$ about a circular bubble.}
\label{fig:schematic}
\end{figure}

Following \cite{Dubash2004,pourzahedi2021}, the critical yield number can be computed directly from:
\begin{equation}\label{eq:Yc}
Y_c ~\equiv \sup_{\boldsymbol{v} \in \mathcal{V},~\boldsymbol{v} \neq 0} \left\{ - \frac{\displaystyle \int_{\Omega \setminus \bar{X}} \boldsymbol{v} \boldsymbol{\cdot} \boldsymbol{e}_g ~\text{d} A}{\displaystyle \int_{\Omega \setminus \bar{X}} \Vert \dot{\ubgamma} \left( \boldsymbol{v} \right) \Vert ~\text{d} A} - \frac{\displaystyle \int_{\partial X} \frac{\gamma}{\kappa} \left( \boldsymbol{v} \boldsymbol{\cdot} \boldsymbol{n} \right) ~\text{d}S}{\displaystyle \int_{\Omega \setminus \bar{X}} \Vert \dot{\ubgamma} \left( \boldsymbol{v} \right) \Vert ~\text{d} A} \right\}
\end{equation}
where $\mathcal{V}$ is the set of admissible velocity fields. As discussed by \citet{pourzahedi2021} for a circular bubble, the surface tension does not change the critical yield number since $\gamma/\kappa$ is constant over $\partial X$ and the flow is divergence free, hence the numerator of the second term in (\ref{eq:Yc}) vanishes. In other words, since the bubble is circular, it is in its equilibrium shape and the only yielding contribution comes from the bubble buoyancy. Hence in the present study, assuming the cloud of bubbles consists of circular bubbles, we neglect the surface tension in what follows.

Our objective is to compute $Y_c(\phi)$ in a meaningful way. \cite{pourzahedi2021} have shown that for a single circular bubble the critical yield number is $Y_{c,0}=0.1718$ (see figure \ref{fig:schematic}); using both the method of characteristics for a perfectly-plastic medium (panel (a)) and computationally using an adaptive augmented Lagrangian scheme (panel (b)). The critical yield number of an individual bubble is the limit of zero volume fraction, i.e.~$Y_{c,0}=Y_c (\phi \to 0)$.

\subsection{Methodology}
\label{sec:methods}

We perform computations with randomized positions of the circular bubbles in a full periodic square box; size of which is $20 \times 20$ (due to scaling the bubble radii $=1$). We handle the bubbles in the numerical simulation by the same method discussed in detail by \cite{pourzahedi2021}. In overview, we use the augmented Lagrangian method coupled with an adaptive finite element method \citep{roquet2003adaptive} implemented in FreeFem++ \citep{MR3043640} to solve equations (\ref{eq:non-govern}) to (\ref{eq:YL_normal}). The computational procedure has been validated extensively in our previous studies \citep{chaparian2017yield,chaparian2020porous,chaparian2020sliding,pourzahedi2021}, and the mesh refinement nicely captures the yield surfaces.

\begin{figure}
\centerline{\includegraphics[width=0.75\linewidth]{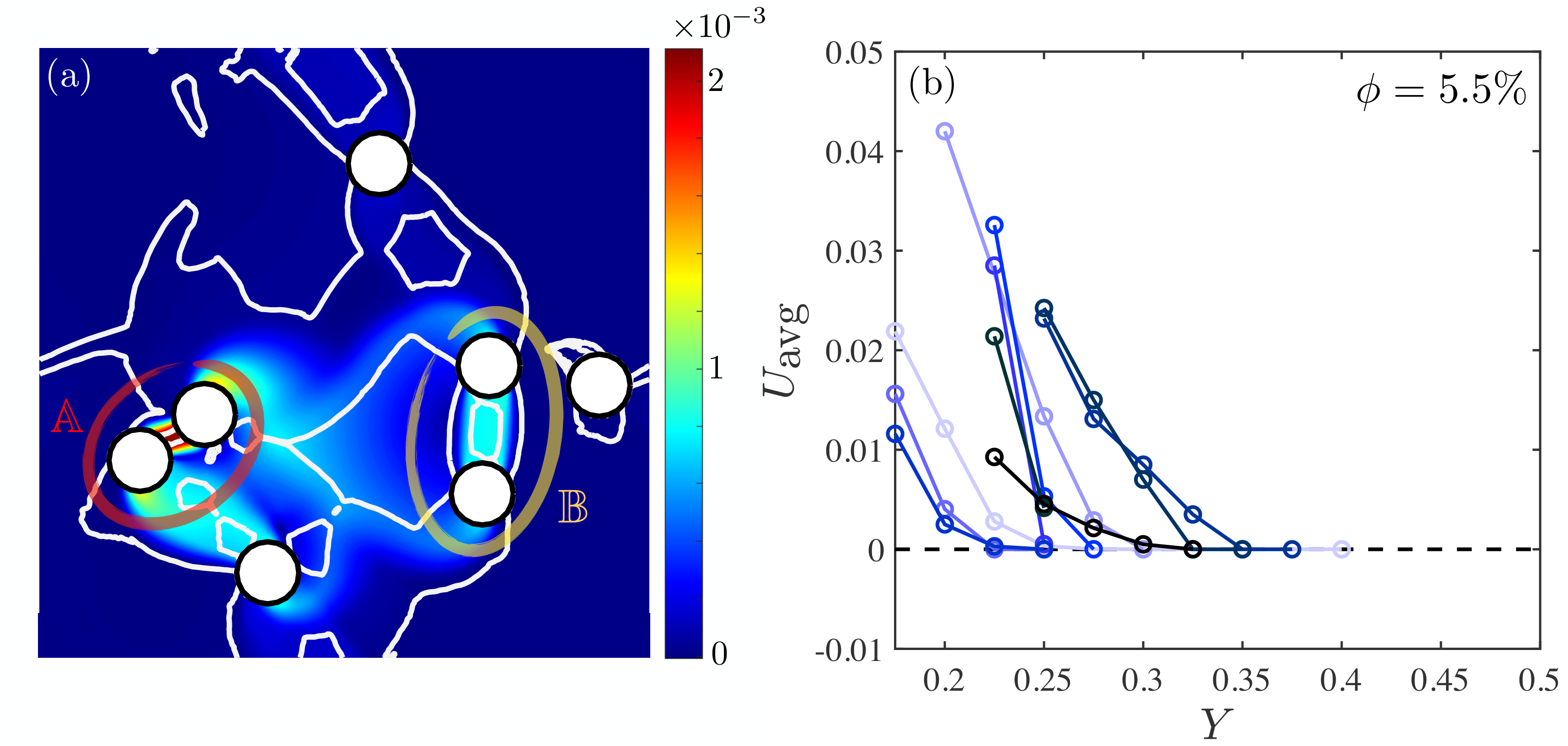}}
\caption{(a) Velocity contour at $Y=0.25$ for a randomized cloud with $\phi=0.055$ (b) Average velocity of the bubbles versus the yield number for different realizations (in blue colour with different intensities). Only a few configurations have been shown to avoid cluttering the figure.}
\label{fig:velocity}
\end{figure}

The numerical experiment protocol is as follows. For a fixed area fraction, based on the size of the computational domain, we calculate the number of bubbles $N$ (i.e.~$\phi=N \pi/L^2$, with here $L=20$) and randomly assign bubble position. By changing the yield number, we assess the average velocity of the bubbles as a function of $Y$ via:
\begin{equation}
N \pi U_{\text{avg}} = \int_{\Omega \setminus \bar{X}} \boldsymbol{u} \cdot \boldsymbol{e}_g ~\text{d}A,
\end{equation}
which follows from the continuity equation, i.e.~what flows up must flow down. A sample computation at $\phi=0.055$ and $Y=0.25$ is shown in figure \ref{fig:velocity}(a).

Having computed $U_{\text{avg}}$ as a function of the yield number (one blue curve in figure \ref{fig:velocity}(b)), we calculate the $Y$ for which the flow stops and hence the critical yield number for one configuration $Y_c^i$. We repeat this procedure for other randomized configurations at the same volume fraction. After computing a large number of different configurations we average the data to approximate $Y_c$ for a specific volume fraction: $ Y_c(\phi) = (\sum_{i=1}^n Y_c^i)/n$. Note that each instance of $Y$ for each configuration requires 4-5 mesh adaptations. Thus the entire calculation is intensive. We ensure that the number of configurations $n$ is enough to reach statistically converged results for the mean (typically $n = 25 $). We also compute the standard deviation of each sequence of $n$ configurations.

\section{Results}\label{sec:results}

\begin{figure}
\centerline{\includegraphics[width=0.5\linewidth]{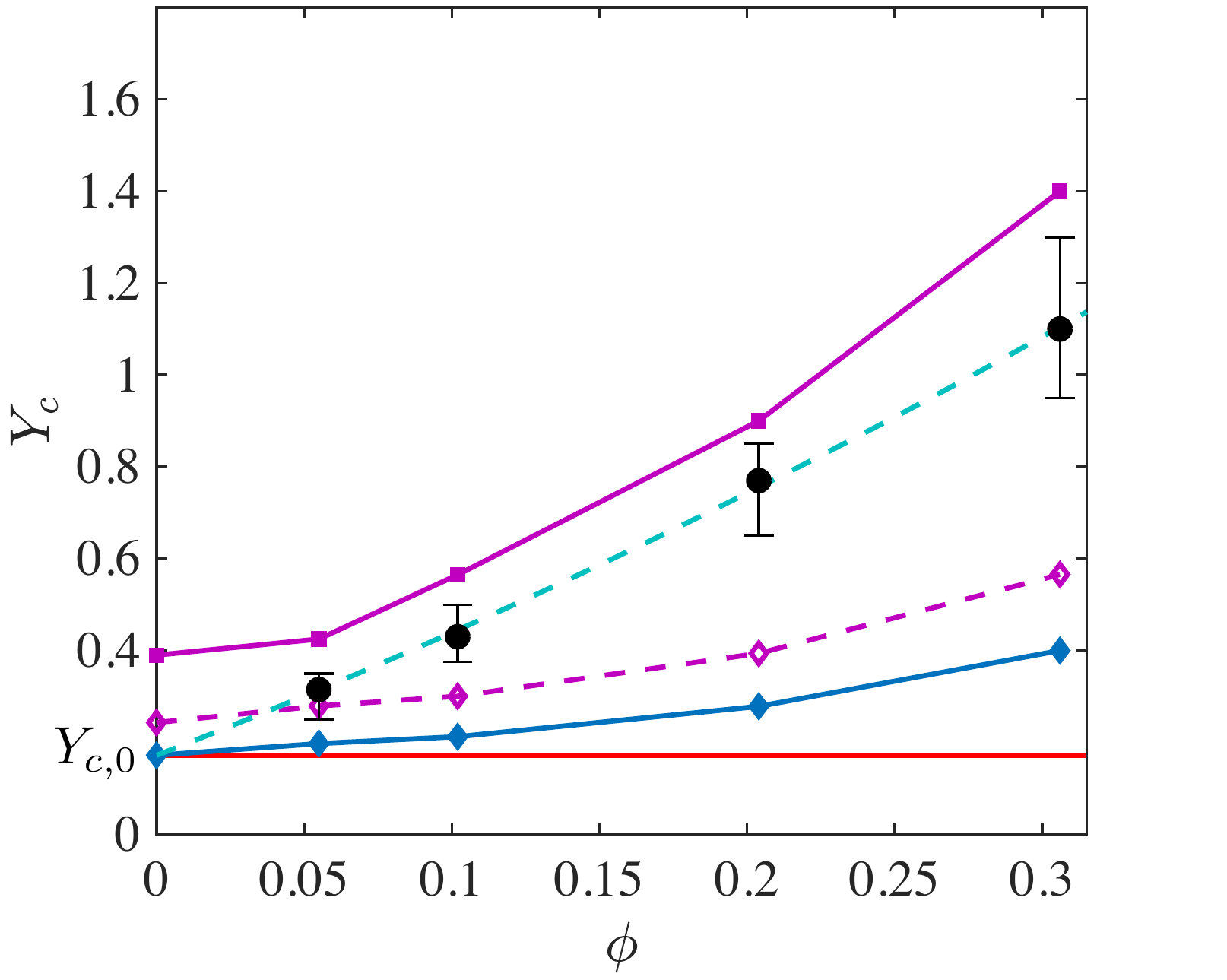}}
\caption{Critical yield number versus bubble concentration $\phi$. The black circle symbols are the randomized cloud bubble simulations results with error bars. The cyan curve is the expression (\ref{eq:Yc_phi_simplemodel1}) \& (\ref{eq:Yc_phi_simplemodel2}), fitted to the data for $0 \leqslant \phi \leqslant 0.31$. The red line marks the single bubble dilute limit: $Y_{c,0} = 0.172$. The blue line shows the results obtained from the equally spaced bubbles, as in figure \ref{fig:periodic}. The purple continuous line shows the results obtained from the equally spaced pairs illustrated in figure \ref{fig:periodic2}. The dashed purple line shows equally spaced large bubbles of equivalent area to the bubble pairs.}
\label{fig:Yc}
\end{figure}

Following the Monte Carlo procedure described above, the computed critical yield number $Y_c(\phi)$ is shown in figure \ref{fig:Yc}, represented by the black circles. The error bars mark the minimum and maximum $Y_c^i$ obtained in the series of randomized configurations at fixed volume fraction.

As depicted, the critical yield number increases with the gas volume fraction, which is intuitive. A similar increase has been shown for non-colloidal particle suspensions recently by \cite{koblitz2018DNS}.
This increase has two main reasons. Firstly, when the amount of gas increases, a larger yield stress is required to stabilize the mixture. Secondly, as demonstrated by \cite{chaparian2018inline}, some networks/clusters of particles can be formed by unyielded bridges which increase $Y_c$ since it is no longer individual bubbles/particles that should be brought to a halt by the yield stress; indeed it is the larger bubbles/particles networks that are the last to stop as $Y$ is increased.

The increase in $Y_c$ is linear at low volume fractions, but clearly deviates from linear behaviour at larger $\phi$. In our methodology, we have increased $Y$ for each configuration until the flow is arrested. The critical yield number is thus the ratio of the critical yield stress to the buoyancy stress. We can represent this as $\hat{\tau}_{y,c}(\phi)$:
\begin{equation}\label{eq:Yc_phi_simplemodel1}
Y_c \left( \phi \right) = \frac{\hat{\tau}_{y,c} \left( \phi \right)}{\Delta \hat{\rho} \hat{g} \hat{R}} = \frac{\hat{\tau}_{y,c} (0)}{\Delta \hat{\rho} \hat{g} \hat{R}} \frac{\hat{\tau}_{y,c} \left( \phi \right)}{\hat{\tau}_{y,c} (0)} = Y_{c,0} \frac{\hat{\tau}_{y,c} \left( \phi \right)}{\hat{\tau}_{y,c} (0)} = Y_{c,0} ~ f(\phi) .
\end{equation}
Here $f(\phi)$ represents the increase in $Y_c(\phi)$ over the single bubble $Y_{c,0}$. On fitting to the data we find:
\begin{equation}\label{eq:Yc_phi_simplemodel2}
f(\phi) = \frac{\hat{\tau}_{y,c} \left( \phi \right)}{\hat{\tau}_{y,c} (0)} = 1 + 14.49 ~\phi + 21.26 ~\frac{\phi^2}{2},
\end{equation}
which is sketched by the broken cyan curve in figure \ref{fig:Yc}. For future reference, the computed data are given in table \ref{tab:Yc}.

\begin{table}
  \begin{center}
\def~{\hphantom{0}}
  \begin{tabular}{cccccc}
    $\phi$    & $n$  &  $Y_c$ & $SD$ & $\min \{ Y_c^i \}$  & $\max \{ Y_c^i \}$  \\[3pt]
       0.055 & 20 & 0.305 & 0.0357 & 0.25 & 0.35 \\
       0.102   & 25 & 0.430 & 0.0525 & 0.375 & 0.5 \\
       0.204   & 30 & 0.769 & 0.1063 & 0.65 & 0.85\\
       0.306   & 30 & 1.110 &  0.1432 & 0.95 & 1.3 \\
  \end{tabular}
  \caption{$Y_c$ and statistics of randomized simulations}
  \label{tab:Yc}
  \end{center}
\end{table}

\subsection{Further analysis and bounds}

\begin{figure}
\centerline{\includegraphics[width=0.6\linewidth]{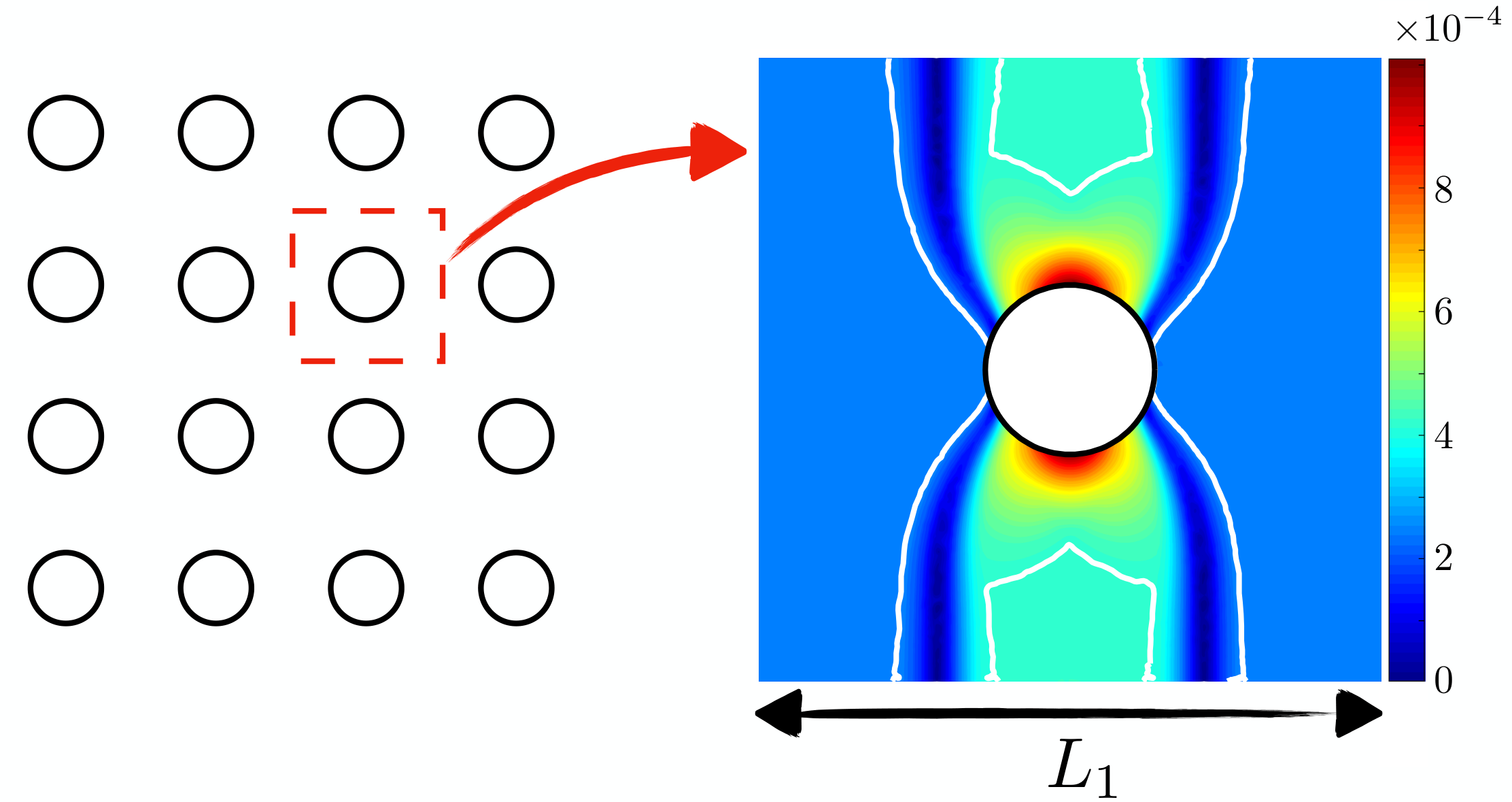}}
\caption{Schematic of the equally spaced bubbles with the desired volume fraction and the sample computation at $Y=0.18$ with $\phi=0.055$ (i.e.~$L_1=7.56$ given that $R=1$).}
\label{fig:periodic}
\end{figure}

Figure \ref{fig:Yc} contains other curves that  shed light on different contributions to the buoyancy-yield stress balance. To get an estimation of the minimal increase in the critical yield number by the increased volume fraction, we simulate the flow around an individual bubble in a periodic box of size $L_1=\sqrt{\pi / \phi}$; see figure \ref{fig:periodic}. In other words, in this simulation we focus on a bubble suspension in which the bubbles are equally spaced and so the hydrodynamic interactions are minimal compared to the randomized bubble cloud. There are other regular spacings (e.g.~hexagonal), but it is reasonable to assume that the in-line arrangement is more likely to yield to motion. The critical yield numbers predicted by these ``conceptual'' suspensions are shown in blue in figure \ref{fig:Yc}. While, as expected, $Y_c$ increases with $\phi$ in these simulations as well, the large gap between the blue line and the circle symbols (cloud data) explicitly demonstrates that short range interactions between the bubbles play an important role in yielding.

\begin{figure}
\centerline{\includegraphics[width=0.9\linewidth]{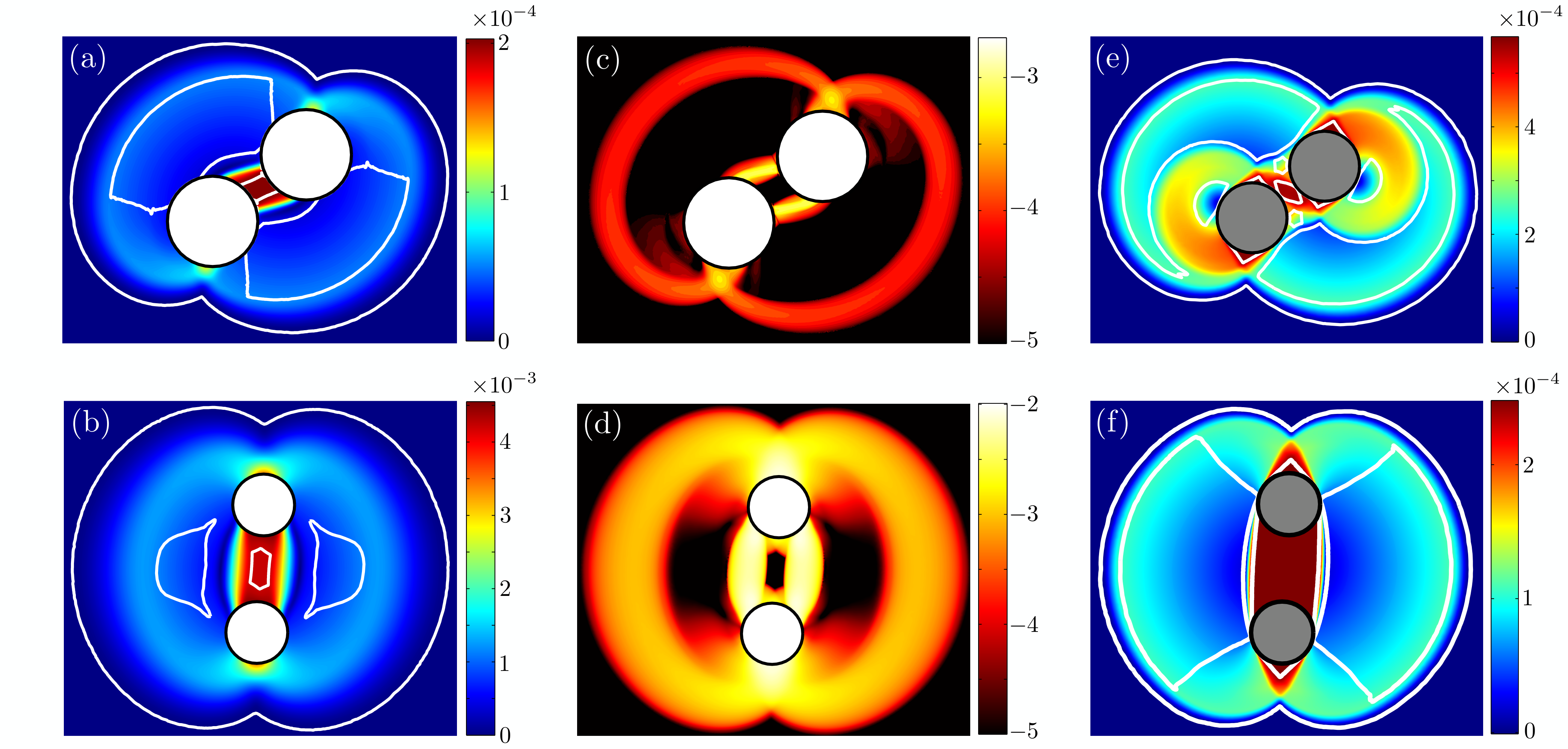}}
\caption{Flow fields around isolated pairs extracted from figure \ref{fig:velocity}: top row panels are associated with the closest pair (shown by red marker in figure \ref{fig:velocity}) and the second closest pair (shown by yellow marker in figure \ref{fig:velocity}). (a,b) Velocity contours for bubbles; (c,d) Contour $\log (\Vert \dot{\ubgamma} \Vert)$ for bubbles; (e,f) Velocity contours around particles. (a,c) $Y=0.225$; (b,d) $Y=0.2$; (e) $Y=0.1425$; (f) $Y=0.17$. Please note that the critical yield number for these configurations are : (a,c) $Y_c=0.25$; (b,d) $Y_c=0.225$; (e) $Y_c=0.145$; (f) $Y_c = 0.1725$.}
\label{fig:comparison}
\end{figure}

For deeper understanding of the short range interactions, we have revisited a couple of cloud simulations in the dilute regime (mostly at $\phi=5.5\%$). We have found that the critical yield number for the cloud is quite close to $Y_c$ of the dominant pair. For instance, for the cloud shown in figure \ref{fig:velocity}(a), the dominant pair is highlighted in red (pair $\mathbb{A}$). It is apparent from the velocity contour that the maximum velocity occurs between these bubbles and this pair is connected by an unyielded bridge. The second dominant pair is highlighted yellow (pair $\mathbb{B}$). We perform simulations in which we just model these pairs ignoring all other bubbles in the cloud and setting  $\mathbf{u}=0$ in the far field. In other words, we simulate the two bubbles which are proximate in an ambient quiescent pool of viscoplastic fluid. Figure \ref{fig:comparison}(a-d) reveals more flow features (velocity and $\log(\Vert \dot{\ubgamma} \Vert)$ fields) around these dominant pairs. The top panels are associated with the pair $\mathbb{A}$ and the bottom panels with the pair $\mathbb{B}$ extracted from the sample simulation shown in figure \ref{fig:velocity}.

The critical yield number for the cloud shown in figure \ref{fig:velocity}(a) is $Y_c=0.265$, for the dominant pair (i.e.~pair $\mathbb{A}$) it is $Y_c=0.25$, and for the second dominant pair (i.e.~pair $\mathbb{B}$) we find $Y_c=0.225$. For the sake of conciseness, we do not compare $Y_c$ of all the simulated clouds with the dominant pair, but in almost all the cases we have checked the two critical yield numbers are approximately the same in the dilute regime.

It should be mentioned that generally finding the dominant pair is not trivial and one can easily imagine cases of non-uniqueness or where a larger cluster is dominant. Nor is the dominant pair necessarily the same for bubbles as for solid particles. For instance, figure \ref{fig:comparison}(e,f) shows the same arrangements of the pairs ($\mathbb{A}$ and $\mathbb{B}$) when they are solid particles. Interestingly, pair $\mathbb{B}$ is the dominant pair in the case of solid particles. Pair $\mathbb{B}$ are almost vertically aligned and this triggers the formation of a unyielded bridge between the solid particles which connects the particles together and increases $Y_c$. More precisely, in the case of bubbles, the larger critical yield number is associated to pair $\mathbb{A}$ ($Y_c=0.25$) whereas in the case of solid particles, the larger critical yield number is associated to pair $\mathbb{B}$ ($Y_c=0.1725$). Hence, in different physical problems, the dominant pair could have different configurations. Indeed, it is a multi-dimensional problem in which proximity and orientation of bubbles/particles are two important parameters.

\begin{figure}
\centerline{\includegraphics[width=0.6\linewidth]{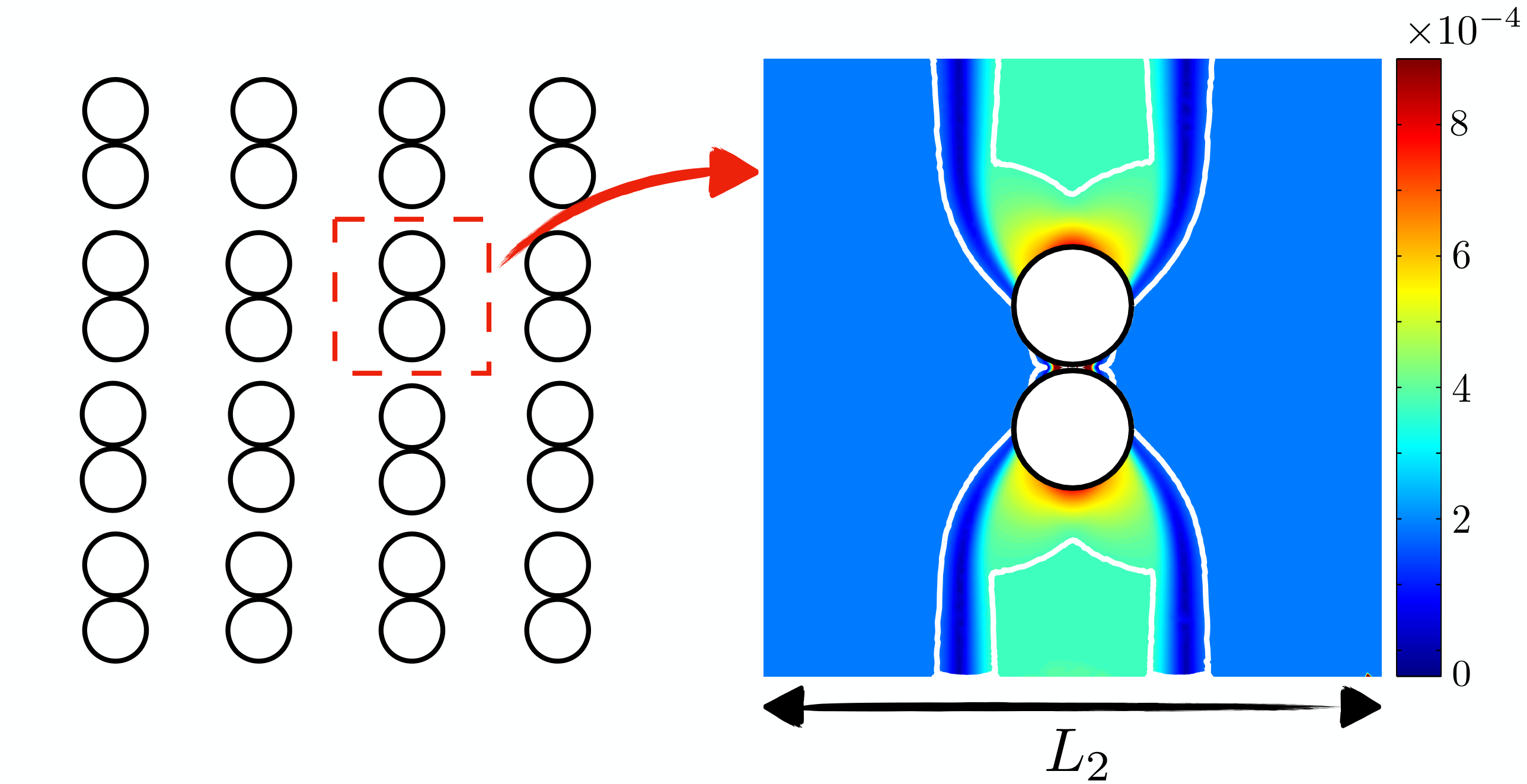}}
\caption{Schematic of the equally spaced twin pairs with the desired volume fraction and the sample computation at $Y=0.4$ with $\phi=0.055$ (i.e.~$L_2=10.69$ given that $R=1$).}
\label{fig:periodic2}
\end{figure}

At higher volume fractions, the whole cloud cannot be reduced to a dominant pair. It is indeed a network of bubbles that controls yielding and  extracting that cluster from a fully packed realization is not trivial. However, to get an estimation, we also investigate another ``designed" suspension in which we force each two bubble pair to have strong short range interaction by almost touching each other when they are aligned vertically; see figure \ref{fig:periodic2}. We again perform simulations in a small periodic box of size $L_2 = \sqrt{2 \pi/\phi}$. The critical yield number of these type of clouds is shown by the purple curve in figure \ref{fig:Yc}. As we see, this leads to an upper bound for the randomized cloud data since the interactions are forcefully increased. However, if the two touching bubbles are merged to form a larger single bubble of equivalent area, the critical yield number is the dashed purple curve in figure \ref{fig:Yc} which gives a much smaller $Y_c$ because the interactions are absent. It is interesting to note that in this sense bubble coalescence may not be optimal for (onset of) motion! This same procedure could be extended by making the interactions even more dramatic such as having a vertical chain of three or four touching bubbles instead of two bubbles, presumably with larger upper bounds for $Y_c$.

\section{Summary \& conclusions}\label{sec:conclusion}

In this study we have focused on clouds of bubbles in a yield-stress fluid and mainly have discussed the static stability of these bubbles. The main objective is to respond to practical problems of  environmental or industrial nature: how much gas fraction can be held in a yield-stress fluid? To this end, we performed exhaustive sets of computations with randomized positions of bubbles in a full periodic box and monitored the average velocity of the bubbles as a function of the yield number $Y$ (i.e.~the ratio of the fluid yield stress to the buoyancy stress). The critical yield number which marks the flow/no flow limit was then extracted  for each bubble cloud and a Monte Carlo procedure was used to determine averaged $Y_c$ as a function of gas fraction.

As expected, we found that for larger volume fractions, the critical yield number is larger. In the dilute regime the behaviour is linear, but for larger volume fractions the increase is more dramatic. To highlight the different contributions, we also performed simulations for equally spaced suspensions of bubbles, which gives a lower bound to $Y_c$ due to the larger gas fraction. Computations for vertically aligned twin touching bubbles lead to an upper bound. The short range interaction of bubbles significantly increases the critical yield number, similar to the formation of clusters in suspension of particles in a yield-stress fluid \citep{chaparian2018inline,koblitz2018DNS}. This fact highlights the importance of computing randomized configurations.

The relevance of randomized distributions is very problem dependent. In situations where bubbles nucleate within a static fluid, e.g.~the oil sands tailing pond application introduced earlier, this is likely reasonable, although mono-sized bubbles are an approximation. Equally, the sensitivity to clustering at higher concentrations is hard to account for, e.g.~it may occur due to initial non-uniformity in naptha concentration. Other bubbly (yield stress) liquids may be more structured e.g.~in a processing flows. Vigorous shaking of bubbly mixtures can also easily result in non-spherical static bubbles, e.g.~see the images in \cite{Dubash2004}. Thus, we are only scratching the surface here. Our work can be, for example, extended to bidispersed/more realistic bubble clouds and also larger $\phi$. Another challenge would be to study foams as a limit where surface tension forces become dominant over the yield stress of the bulk.

Our study opens new perspectives in the study of bubbly flows of yield-stress fluids and even more complex multiphase systems of gels and pastes. In recent years, the knowledge of particle/bubble suspensions in yield-stress fluids has mostly expanded in the rheological studies \citep{dagois2015rheology,kogan2013mixtures}, i.e.~for a given volume fraction how is the bulk rheology of the mixture changed? Typically this results in a multiplicative scaling of the rheological constants.
Here we too have such a scaling, captured in $f(\phi)$; see expression (\ref{eq:Yc_phi_simplemodel2}). Note that the linear increase in $f(\phi)$ is much larger than those of rheological closures. The point to emphasize is that there are 2 quite different considerations: (i) the rheology of a bubbly mixture (with no density difference between phases) when placed under shear, extension etc., and (ii) the limit under which buoyancy driven bubble flows do not occur: studied here for the first time.

\section*{Acknowledgements}

This research was made possible by collaborative research funding from NSERC and COSIA/IOSI (project numbers CRDPJ 537806-18 and IOSI Project 2018-10). This funding is gratefully acknowledged. This computational research was also partly enabled by infrastructure provided from Compute Canada/Calcul Canada (www.computecanada.ca).

\section*{Declaration of Interests}

The authors report no conflicts of interest.

\bibliographystyle{jfm}

\bibliography{viscoplastic}


%
%
%
%
%
%
%


\end{document}